# Resonant multi-dielectric coverslip for enhanced total internal reflection fluorescence microscopy

Y. Toumi[1], A. Mouttou[1,2], F. Lemarchand[1], G. Demesy[1], C. Koc[1], D. Muriaux[2], A. Moreau[1],
J. Lumeau[1], C. Favard[2,**], A. L. Lereu[1,*]

[1] *Aix Marseille Univ, CNRS, Centrale Marseille, Institut Fresnel, Marseille, France*

[2] *Institut de Recherche en Infectiologie de Montpellier, CNRS, Univ of Montpellier France*

[*] aude.lereu@fresnel.fr

[**] cyril.favard@irim.cnrs.fr

**ABSTRACT**

Multi-dielectric coated coverslip can be designed to reach large optical field enhancements when working under proper illumination conditions and in total internal reflection. In an objective-based total internal reflection fluorescence microscopy (TIRF-M), we propose to use the resulting large field enhancement supported in such coverslip to improve TIRF-M sensitivity by amplifying the collected fluorescence signal. We present here the optimization, realization and testing in TIRF-M of resonant coverslips designed to take into account the experimental constraints. The latter's are due to the inverted configuration of TIRF-M together with the use of a high numerical aperture objective. The challenge is therefore to find the best coating design compromising between the angular tolerance and the field enhancement. We will report here enhanced-TIRF-M imaging over model samples but the ultimate goal here is to be able to monitor and quantify dynamics of pathogens at the membrane of living cells.

## 1. INTRODUCTION

Total Internal Reflection Fluorescence microscopy also called TIRF microscopy is a technique that uses evanescent wave illumination to selectively excite fluorophores in a thin layer of the sample close to the coverslip. TIRF microscopy is particularly useful for studying processes that occur at the cell membrane. This technique, as its name suggests, is based on total reflection which occurs when the incident angle is greater than the critical angle. However, TIRF microscopy suffers from various limitations which can affect the optical contrast and the lateral resolution of the obtained biological images. For example, the problem of uneven illumination, due to the beam divergence, can degrade the image quality. In fact, the evanescent wave at the interface between the coverslip and the biological sample reaches its maximum intensity when the sample is illuminated at the critical angle. Yet, the exciting beam is inherently divergent thus a part of the beam is totally reflected which leads to the generation of the evanescent beam and another part, i.e the rays coming at angles less than the critical angle, propagates through the sample. To overcome this issue, we could increase the incident angle to guarantee a pure evanescent illumination of the sample. Unfortunately, this leads to a decay of the intensity of the evanescent field which degrades again the image quality. Dielectric thin films suggest promising prospects for amplifying the sensitivity of TIRF imaging due to their extreme confinement field property at the free interface when properly designed. Thin films are thin layers of material with a thickness typically ranging from a few nanometers to a few micrometers. They are used to modify the behavior of light, by reflecting, transmitting, or absorbing it in specific ways. Hence dielectric multilayers (DM) can be used to enhance locally the evanescent wave without introducing any further modifications into the optical arrangement of a traditional TIRF microscope.

We showed that DM can be designed to support a desired field enhancement regardless of the excitation wavelength and incident angle [1], [2], [3] and with a defined enhancement at the free interface; here the biological medium. In the context of TIRF microscopy, we demonstrated the critical role of the angular divergence over the resulting fluorescence enhancement and the impact of the DM transmission over the spectral and angular range of the fluorescence collection [4]. Here we discuss a last parameter to be taken into account in the optimization of the DM to have a good theory-experiment agreement when measuring the fluorescence from beads [5].

## 2. RESULTS AND DISCUSSIONS

In this work, the DM stack consists of alternating quarter wave layers of $SiO_2$, the low refractive index material, and $Nb_2O_5$, the high refractive index material. The stack ends with a layer of $SiO_2$ to guarantee the biocompatibility. A stack is optimized to be resonant at an excitation wavelength of 561 nm and at an incident angle of 68° by adjusting the number of layers and the various thicknesses. Furthermore, the last layer of $SiO_2$ was replaced by a layer of $SiO_i$, i < 2 to control the absorption and therefore adapt the field enhancement and the resonance angular tolerance. In fact, the extinction coefficient *k* of the top layer is a key parameter to control the field enhancement. However, classically it is difficult to experimentally determined values of $k < 10^{-4} - 10^{-5}$, which is the typical value of dielectric thin film. Therefore, to fully control the optimization and obtain a good agreement between the predicted and experimentally measured fluorescence enhancement, we choose to use a material with a measurable $k$ for the top layer of the DM. In order to preserve the bio compatibility of $SiO_2$, we used $SiO_i$, i < 2 a derivative of $SiO_2$ by acting on the oxidization level of the Si target during deposition. We have realized and characterized by spectrophotometry a monolayer of $SiO_x$. The index of this layer was measured to be ñ (561 nm) = $1.602 + i3.2 \ 10^{-3}$.

The DM structure was optimized to be resonant at the same wavelength and incident angle as the initial stack. Therefore the physical thicknesses of the different layers of the stack were adjusted to be optimal at the operating point fixed by the biological application as shown in Fig. 1(a).

A numerical evaluation was then done to study the effect of a variation of the $k$ value of the $SiO_i$ layer on the fluorescence signal. The fluorescence enhancement factor was estimated for different values of $k$ for the top layer, ranging from $10^{-5}$ to $10^{-2}$ while keeping the same physical thickness. An increase in the fluorescence enhancement was expected while decreasing $k$ until than it reaches a plateau for $k < 10^{-4}$.

Experimentallly, three DM coated coverslips were made for this study; one ending with a layer of $SiO_2$, one with a layer of $SiO_x$ and one with $SiO_y$ with x < y < 2.

The fluorescence enhancement factor was introduced in a previous work [4] as the product of the excitation enhancement and the DM average transmission $Tav$ over the spectral and angular range of the fluorescence collection. The excitation enhancement, named $\xi(\Delta\theta, \theta)$, is defined by the ratio of the intensity measured at the DM-coated coverslip free interface to the one measured at the glass coverslip free interface, at the resonance angle $\theta$ and for a beam divergence $\Delta\theta$. The expected fluorescence enhancement factor $\xi(\Delta\theta, \theta).Tav$ was then calculated as a function of incident angle for the three considered DM and for a classical glass coverslip as reference, see in Fig. 1(b). The maxima were obtained at the expected angular positions. The fluorescence enhancements were observed to be ×2.91, ×3.45 and ×4.86 with respect to the glass coverslip, for DM ($SiO_x$), DM ($SiO_y$) and DM ($SiO_2$), respectively. Together with the field enhancement variation, a change in the angular tolerance for the three considered DM resonance was also observed. In fact, when the field enhancement increases, the full width at half maximum (FWHM) of the resonance decreases, in the presented cases, from 0.16° for DM ($SiO_x$), to 0.13° for DM ($SiO_y$) and even 0.038° for DM ($SiO_2$). Or, for the fluorescence enhancement in an objective-based TIRF microscope, we need to find the best compromise between large field enhancement and angular tolerance to achieve the best fluorescence enhancement.

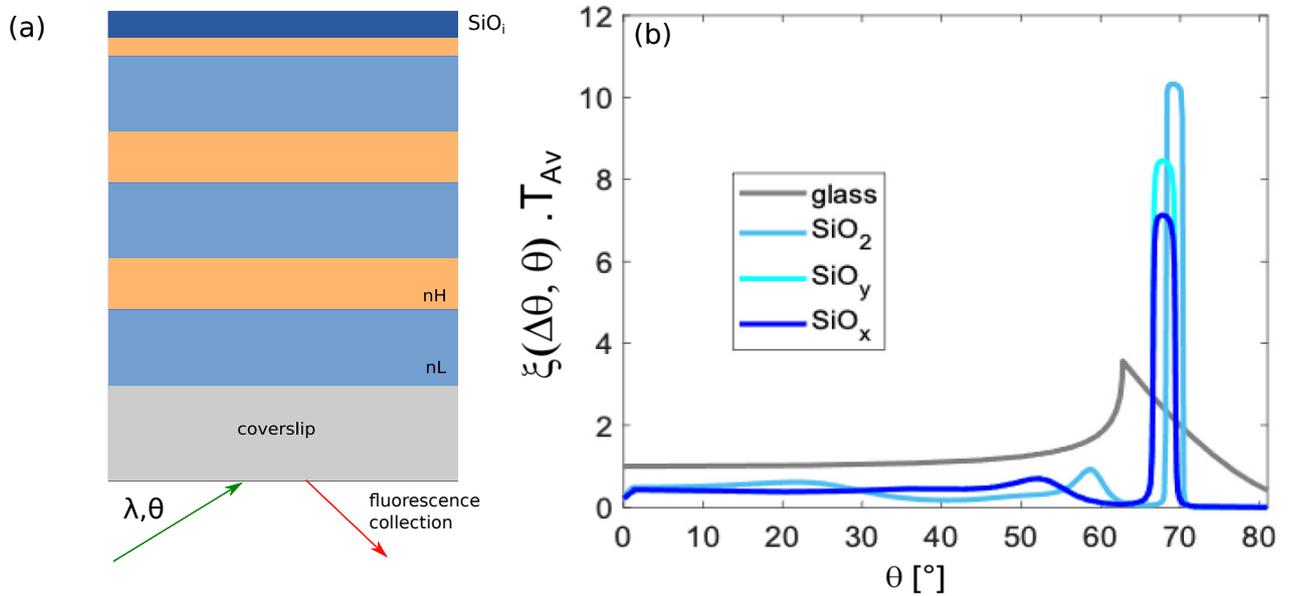

Figure 1. (*a*): DM($SiO_x$) stack with an absorbing top layer of SiOx of extinction coefficient $k=3.2 \ 10^{-3}$
(*b*): The angular distribution of the fluorescence enhancement factor $\xi(\Delta\theta, \theta).Tav$, for the three samples (blue curves) compared to classical glass coverslip (gray curve) [2].

### 3. CONCLUSION

In conclusion, resonant dielectric multilayer stacks were designed to enhance the evanescent field at the excitation with an angular tolerance in accordance with the angular divergence fixed by the commercial TIRF-

M. The DM transmission efficiency, over the angular and spectral range of the fluorescence collection, was also considered to compare the expected fluorescence with the measured one. This leads to improving the sensitivity of TIRF Microscopy by a factor of 4 for an angular divergence of 10 mrad. Here we will present the numerical predictions and the associated experiment in TIRF microscopy over fluorescence beads.

## ACKNOWLEDGEMENTS

The authors acknowledge the CNRS for financial supports through the 80|PRIME interdisciplinary program and the French research agency through the ANR NIS.

## REFERENCES

[1] C. Ndiaye, F. Lemarchand, M. Zerrad, D. Ausserré, and C. Amra: Optimal design for 100% absorption and maximum field enhancement in thin-film multilayers at resonances under total reflection, Appl. Opt. 50(9), C382, 2011.
[2] A. L. Lereu, M. Zerrad, M. Petit, F. De Fornel, and C. Amra: Multi-dielectric stacks as a platform for giant optical field, Proc. SPIE 9162, 916219, 2014.
[3] A. L. Lereu, M. Zerrad, C. Ndiaye, F. Lemarchand, and C. Amra: Scattering losses in multidielectric structures designed for giant optical field enhancement, Appl. Opt. 53(4), A412, 2014.
[4] A. Mouttou, F. Lemarchand, C. Koc, A.Moreau, J. Lumeau, C. Favard, and A. L Lereu: Resonant dielectric multilayer with controlled absorption for enhanced total internal reflection fluorescence microscopy, Optics Express, 30(9):15365–15375, 2022.
[5] A. Mouttou, F. Lemarchand, C. Koc, A. Moreau, J. Lumeau, C. Favard, and A. L Lereu: Optimization of resonant dielectric multilayer for enhanced fluorescence imaging, Optical Materials: X, 17:100223, 2023.